\documentclass[%
superscriptaddress,
 amsmath,amssymb,
floatfix,
twocolumn
]{revtex4-2}

\usepackage[utf8]{inputenc}
\usepackage[T1]{fontenc}
\usepackage{etoolbox}

\usepackage[dvipdfm]{graphicx} 
\usepackage{amsmath}
\usepackage{physics}
\usepackage{xcolor}
\usepackage{bm}
\usepackage{mathtools}
\usepackage[ruled,lined]{algorithm2e}
\usepackage{chemformula}
\usepackage{titlesec}
\usepackage[acronym]{glossaries}
\usepackage{enumitem}
\usepackage{grffile}
\usepackage{listings}

\definecolor{codegreen}{rgb}{0,0.6,0}
\definecolor{codegray}{rgb}{0.5,0.5,0.5}
\definecolor{codepurple}{rgb}{0.58,0,0.82}
\definecolor{backcolour}{rgb}{0.95,0.95,0.92}

\lstdefinestyle{mystyle}{
    backgroundcolor=\color{backcolour},   
    commentstyle=\color{codegreen},
    keywordstyle=\color{magenta},
    numberstyle=\tiny\color{codegray},
    stringstyle=\color{codepurple},
    basicstyle=\ttfamily\footnotesize,
    breakatwhitespace=false,         
    breaklines=true,                 
    captionpos=b,                    
    keepspaces=true,                 
    numbers=left,                    
    numbersep=5pt,                  
    showspaces=false,                
    showstringspaces=false,
    showtabs=false,                  
    tabsize=2
}

\newacronym{ml}{ML}{machine learning}
\newacronym{mlff}{ML-FF}{machine learning force field}
\newacronym{qm}{QM}{quantum mechanical}
\newacronym{pso}{PSO}{Particle Swarm Optimisation}
\newacronym{rbf}{RBF}{radial basis function}
\newacronym{airss}{AIRSS}{\textit{ab initio} random structure search}

\newcommand{\mt}[1]{\mathrm{#1}}
\DeclarePairedDelimiter{\ceil}{\lceil}{\rceil}

\titleformat*{\paragraph}{\normalfont\bfseries}

\begin{document}

\title{Zero Shot Molecular Generation via Similarity Kernels}

\author{Rokas Elijošius}
\affiliation{Engineering Laboratory, University of Cambridge, Trumpington St and JJ Thomson Ave, Cambridge, UK}

\author{Fabian Zills}
\affiliation{Institute for Computational Physics, University of Stuttgart, 70569 Stuttgart, Germany}

\author{Ilyes Batatia}
\affiliation{Engineering Laboratory, University of Cambridge, Trumpington St and JJ Thomson Ave, Cambridge, UK}

\author{Sam Walton Norwood}
\affiliation{Department of Energy Conversion and Storage, Technical University of Denmark, Anker Engelunds Vej 301, 2800 Kgs. Lyngby, Denmark}

\author{D\'avid P\'eter Kov\'acs}
\affiliation{Engineering Laboratory, University of Cambridge, Trumpington St and JJ Thomson Ave, Cambridge, UK}

\author{Christian Holm}
\affiliation{Institute for Computational Physics, University of Stuttgart, 70569 Stuttgart, Germany}

\author{G\'abor Cs\'anyi}
\affiliation{Engineering Laboratory, University of Cambridge, Trumpington St and JJ Thomson Ave, Cambridge, UK}

\onecolumngrid
\begin{abstract}

Generative modelling aims to accelerate the discovery of novel chemicals by directly proposing structures with desirable properties. 
Recently, score-based, or diffusion, generative models have significantly outperformed previous approaches.
Key to their success is the close relationship between the score and physical force, allowing the use of powerful equivariant neural networks.
However, the behaviour of the learnt score is not yet well understood.
Here, we analyse the score by training an energy-based diffusion model for molecular generation.
We find that during the generation the score resembles a restorative potential initially and a quantum-mechanical force at the end.
In between the two endpoints, it exhibits special properties that enable the building of large molecules.
Using insights from the trained model, we present Similarity-based Molecular Generation (SiMGen), a new method for zero shot molecular generation.
SiMGen combines a time-dependent similarity kernel with descriptors from a pretrained machine learning force field to generate molecules without any further training.
Our approach allows full control over the molecular shape through point cloud priors and supports conditional generation.
We also release an interactive web tool that allows users to generate structures with SiMGen online (\url{https://zndraw.icp.uni-stuttgart.de}).

\end{abstract}

\twocolumngrid

\maketitle

\section{Introduction}

The combinatorial scaling of the available chemical space with molecule size is one of the main challenges in the design of new molecules and materials. {\em Generative modelling} aims to solve this by directly proposing structures with desirable properties, without exhaustively enumerating and screening candidates. Recently, diffusion-based models have achieved impressive results in molecular docking~\cite{corsoDiffDockDiffusionSteps2022} and generation of linkers~\cite{igashovEquivariant3DConditionalDiffusion2022}, drug-like molecules~\cite{schneuingStructurebasedDrugDesign2022, linDiffBPGenerativeDiffusion2022} and crystal structures~\cite{xieCrystalDiffusionVariational2022, zeni2023mattergen}.

Diffusion models are trained to reverse a stochastic \textit{noising} process, which gradually corrupts samples of training data until they are indistinguishable from samples drawn from an uninformative prior distribution, such as a standard Gaussian~\cite{hoDenoisingDiffusionProbabilistic2020, sohl-dicksteinDeepUnsupervisedLearning2015, songScoreBasedGenerativeModeling2020}. Once the reverse of this process is learnt, it can be applied to transform easily-sampled random noise into independent samples from the approximate data distribution. In particular, diffusion models regress the \textit{score} $\bm{s}(\bm{x},t)$, defined as the gradient of the log-likelihood of a time-dependent distribution $p(\bm{x};t)$ that interpolates between the data distribution at $t=0$, $p(\bm{x};0)=p_\mt{data}(\bm{x})$, and a Gaussian distribution at $t=T$, $p(\bm{x};T)=N\left(\bm{x};\bm{\mu}=\bm{0}, \bm{\Sigma}=\bm{I}\sigma(T)\right)$.
\begin{equation} \label{eqn:score_definition}
    \bm{s}(\bm{x},t) = \nabla_{\bm{x}}\log {p(\bm{x};t)}
\end{equation}
Using the score, we can generate new samples by numerically integrating a stochastic differential equation describing the time reversal of the noising process, which has a known analytic form~\cite{songScoreBasedGenerativeModeling2020}.
Because the data distribution is provided only in the form of samples, the score is intractable analytically. 
However, it can be efficiently learnt via an implicit objective that has a closed form when the limiting distribution at $t=T$ is taken to be Gaussian, an approach known as \textit{denoising score matching}~\cite{JMLR:v6:hyvarinen05a, vincentConnectionScoreMatching2011, songGenerativeModelingEstimating2020}.
 
In the context of molecule generation, the score is closely related to {\em atomic forces}. Consider training data that comprise configurations sampled using molecular dynamics or other methods from an underlying Boltzmann distribution, $\bm{x}\sim\exp\left(-\beta U(\bm{x})\right) / Z$. Here, $\bm{x}=\{\bm{r},\bm{z}\}$ is a set that represents a molecule, with $\bm{r}$ the atomic positions and $\bm{z}$ the chemical elements, $U(\bm{x})$ the potential energy, $\beta$ the inverse temperature, and $Z$ the partition function. In this case, when the elements $\bm{z}$ are fixed, the score of the data distribution $\bm{s}(\bm{x},0)$ corresponds to the atomic force (defined as the negative gradient of the potential energy) up to a multiplicative constant:
\begin{equation} \label{eqn:score_force_equivalence}
    \bm{s}(\bm{x},0) = \nabla_{\bm{r}}\log p(\bm{x};t=0) = -\beta \nabla_{\bm{r}}U(\bm{x}) = \beta \bm{F}(\bm{x})
\end{equation}
The significance is that the atomistic modelling community has built up an extensive body of knowledge on building excellent models of atomic forces in the form of {\em force fields}~\cite{kohnSelfConsistentEquationsIncluding1965,kresseInitioMolecularDynamics1993, jorgensenDevelopmentTestingOPLS1996, wangDevelopmentTestingGeneral2004,clarkFirstPrinciplesMethods2005, vanommeslaegheCHARMMGeneralForce2010, kuhneCP2KElectronicStructure2020, spicherRobustAtomisticModeling2020}, lessons from building machine learning force fields~\cite{behlerGeneralizedNeuralNetworkRepresentation2007, bartokGaussianApproximationPotentials2010, s.smithANI1ExtensibleNeural2017, drautzAtomicClusterExpansion2019} being particularly relevant.
Several groups have already used this score-force relationship to pre-train networks for molecular property prediction~\cite{zaidiPretrainingDenoisingMolecular2022}, to improve the quality of generated structures~\cite{xieCrystalDiffusionVariational2022, wu2022diffusionbased}, or to obtain coarse-grained force fields~\cite{artsTwoOneDiffusion2023}.

Starting from a random configuration and using a force-like quantity to move atoms and generate new structures closely resembles \gls{airss} pioneered by \citet{pickardInitioRandomStructure2011} for crystal structures. In \gls{airss}, atoms are initialised with random positions and then their energy or enthalpy minimized ({\em relaxed}, in common parlance) using \gls{qm} forces, i.e. $\bm{s}(\bm{x},t)=\bm{F}_{\mt{QM}}(\bm{x})$. Despite its simplicity, \gls{airss} has achieved great success in discovering stable structures of periodic materials~\cite{pickardHighPressurePhasesSilane2006, pengHydrogenClathrateStructures2017}, especially at high pressures. However, it is not directly suitable for generating complex molecules. Applying \gls{airss} for molecular generation samples the \gls{qm} energy surface, in other words, the Boltzmann distribution.
This turns out to be very uninteresting: The probability density for molecular systems is overwhelmingly concentrated on the most stable chemical compounds which are simple molecules such as water, carbon dioxide, and dinitrogen. Generative modelling requires a way to {\em guide} the generation towards molecules of interest, and by replacing the learnt score with analytical \gls{qm} forces, that would be lost. Instead, we need a method that allows such directed generation but at the same time leverages the information represented by the available analytical atomic forces.

A further problem of using \gls{airss} for molecule generation is that moving atoms along atomic forces does not allow for a change in composition. This is not a good restriction to have for generating large molecules, because local composition is closely related to local geometry. To retain compositional degrees of freedom in the generative process, score-based models split the score into two components: a positional component $\bm{s}_{\bm{r}}(\bm{x},t)$ that resembles the physical force and an elemental component $\bm{s}_{\bm{z}}(\bm{x},t)$, best interpreted as an alchemical force,
\begin{equation}\label{eqn:score_pos_and_z}
    \bm{s}(\bm{x},t) = \underbrace{\frac{\delta}{\delta \bm{r}}\log p(\bm{x};t) d\bm{r}}_{\bm{s}_{\bm{r}}(\bm{x},t)} + \underbrace{\frac{\delta}{\delta \bm{z}}\log p(\bm{x};t) d\bm{z}}_{\bm{s}_{\bm{z}}(\bm{x},t)}
\end{equation}
The score of the positions $\bm{s}_{\bm{r}}(\bm{x},t)$ obeys the same symmetries as a physical force; namely, it must be translationally invariant and rotationally equivariant. On the other hand, $\bm{s}_{\bm{z}}(\bm{x},t)$ should be both translationally and rotationally \textit{invariant}. 

Given the symmetry requirements, it is no surprise that most recent works have used equivariant graph neural networks to learn the score~\cite{xuGeoDiffGeometricDiffusion2022, hoogeboomEquivariantDiffusionMolecule2022, huaMUDiffUnifiedDiffusion2023, leNavigatingDesignSpace2023a}. Given the success of these architectures for \gls{qm} force predictions~\cite{batatiaMACEHigherOrder2022, batznerEquivariantGraphNeural2022, satorrasEquivariantGraphNeural2022, andersonCormorantCovariantMolecular2019}, they are well suited for learning scores. A key operational principle of these networks is message passing, wherein the representation of each node in a graph (corresponding to an atom in a molecule) is updated based on the state of its neighbouring nodes. Successive layers of message passing encode increasingly higher body-order information in the node's representations~\cite{batatia2022design}. When building \gls{ml} force fields, locality is typically imposed by only allowing atoms that are within a certain cutoff distance of each other to communicate. This enables scaling to large structures by providing linear computational complexity in the number of atoms, and captures the physical intuition that most interactions are short-ranged. In contrast, all previous work on generative modelling for molecules has employed {\em global} models which represent molecules as fully-connected graphs. This poses two challenges: global models are inherently hard to scale, and the model cannot exploit the inductive bias of locality when learning the force-like positional score $\bm{s}_{\bm{r}}(\bm{x},t)$.

Score-based generative models are usually trained to output the score directly, e.g., by applying a linear projection to the learnt features. An alternative is to enforce the conservation of probability by training an energy-based diffusion model~\cite{gaoLearningEnergyBasedModels2021, salimans2021should}, where the score is fitted as a derivative of the model's output. The energy-based formulation comes with a higher computational cost, requiring gradient backpropagation through the network to make predictions. However, the integral of the score corresponds to a learnt molecular potential energy and enables us to study the ``energy landscape'' of the trained generative model. We find in the following that this surface is surprisingly smooth and inherently penalises fragmentation. We suggest that these properties of the learnt energy together with the alchemical force $\bm{s}_{\bm{z}}(\bm{x},t)$  enable diffusion-based models to build larger molecules, compared to the conceptually related \gls{airss} method, which often generates fragmented molecules.

With this insight into the different behaviour of \gls{qm} force fields and purpose-trained diffusion models, we propose a way to generate molecules without the need to train a specialised generative model at all. Our proposal involves constructing a smooth log-likelihood landscape over pretrained descriptors of local atomic environments. We employ a time-varying local similarity kernel (defined by a user-provided dataset) and the descriptors of a pre-trained MACE \gls{qm} force field model~\cite{kovacsMACEOFF23TransferableMachine2023} to define a score-based generative process. The MACE descriptors are locally defined; thus, the entire generation is also local and extensive, allowing us to construct molecules of arbitrary size, as well as perform conditional generation with strictly no changes to the model. The score model can be controlled without retraining by simply substituting a new reference set of local atomic environments for the similarity kernel. We show how further control over the shape of generated structures can be achieved via a point cloud prior. Combining the locality of the kernel with the shaping of the prior allows us to build linkers as well as macrocycle-like molecules with more than 100 heavy atoms. Finally, we release an interactive web browser based tool that enables users to easily carry out conditional generation, which we showcase by building a linker between two molecular fragments.

\section{Results and discussion}

\subsection{Energy landscape of a diffusion model}\label{section:trained-model-and-airss}

\begin{figure*}[ht!]
    \centering
    \includegraphics[width=0.9\textwidth]{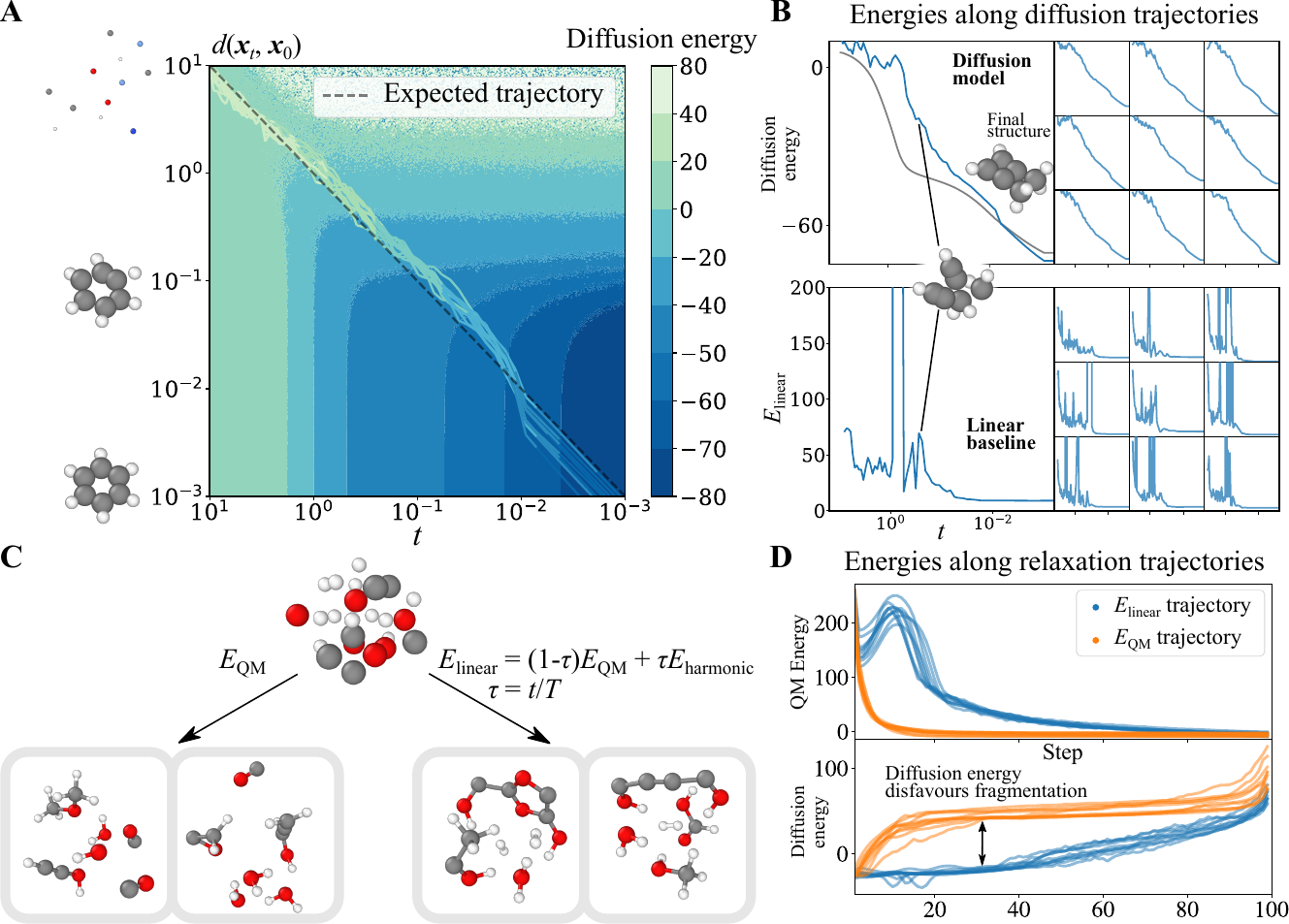}
    \caption{
        \textbf{A. The energy-based diffusion model learns a smooth energy landscape} 
        The landscape depicts the learnt energy of a benzene molecule at different combinations of deformation level (see the main text for explanation) and time. 
        The coloured lines along the diagonal show how the energy and deformation level changes in real generation trajectories. Note that these trajectories do not necessarily result in benzene. However, the energy as a function of time and deformation from the final molecule are similar for all generated molecules and closely mimic the landscape drawn here for benzene.
        \textbf{B. Energy profiles of diffusion trajectories}
        Across ten examples, the learnt energy smoothly deceases during generation, even when the atoms overlap. In contrast, $E_\mt{linear}$ \textit{evaluated on the same trajectories} has a much rougher profile. The grey line shows the energy along an ODE trajectory, while the blue lines correspond to the usual SDE solver.
        \textbf{C. Baseline methods using just QM force fields generate fragmented molecules} Relaxing using just the \gls{qm} forces, i.e. doing \gls{airss} exactly, leads to small fragmented molecules. Relaxation according to $E_\mt{linear}$ produces larger but high-energy molecules.
        \textbf{D. Diffusion models penalise fragmentation} Evaluating a diffusion model on relaxation trajectories reveals that the model has learnt an energy penalty for fragmented structures. For $E_\mt{linear}$ relaxation trajectories, the \gls {qm} energy initially increases as the potential is dominated by the restorative term.
    }
    \label{fig:diffusion_energy_landscape}
\end{figure*}

We start our preliminary investigation by training an energy-based diffusion model, and assume that the time-dependent probability connecting the data distribution to a Gaussian takes the form 
\begin{equation}\label{eqn:energy-probability-relation}
p(\bm{x};t)=\exp\left(- E(\bm{x};t)\right)/Z(t).
\end{equation}
With this formulation, the score is the derivative of a learnt time-dependent energy $E(\bm{x};t)$.
\begin{align}
    \bm{s}_{\bm{r}}(\bm{x},t) &= -\nabla_{\bm{r}}E(\bm{x},t),\\
    \bm{s}_{\bm{z}}(\bm{x},t) &= -\nabla_{\bm{z}}E(\bm{x},t).
\end{align}
Here, $E(\bm{x};t)$ is parametrised using a MACE model, trained using denoising score matching~\cite{JMLR:v6:hyvarinen05a, vincentConnectionScoreMatching2011, songGenerativeModelingEstimating2020}.

Linear interpolation in time between a quadratic potential and the \gls{qm} energy provides a natural baseline to compare the behaviour of the learnt generation process to:
\begin{equation}\label{eqn:linear_pot_interpolation}
E_{\text{linear}}(\bm{x};t) = (1-\frac{t}{T})E_{\text{QM}}(\bm{x}) + \frac{t}{T}||\bm{r}||_2^2/(2\sigma(T)^2)
\end{equation}
This linear baseline is loosely analogous to \gls{airss} with an initially very high pressure, where forces are assumed to be dominated by the confinement, with the pressure decreasing during generation. Like \gls{airss}, it requires a fixed composition and can only sample the global Boltzmann distribution. The time intervals where the learnt energy, $E(\bm{x},t)$, differs most from this linear baseline are thus the most critical. In these regions, the diffusion model has learnt to guide generation towards chemical space similar to the training data, rather than towards the modes of the global Boltzmann distribution.

\begin{figure*}[ht!]
    \centering
    \includegraphics[width=1\textwidth]{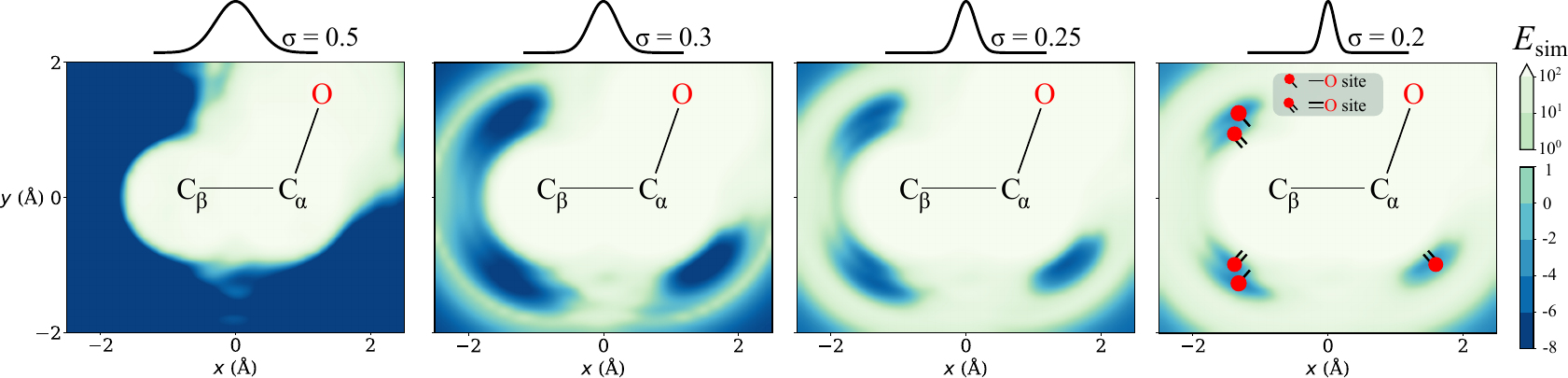}
    \caption{\textbf{The kernel width modulates the similarity energy landscape} Each point in the plots represents $E_\mt{sim}$ if an extra oxygen atom was added at that location. Varying $\sigma(t)$ creates a funnelling effect that pushes atoms from the flat surface at large $\sigma$ to specific minima at the end. }
    \label{fig:effect_of_kernel_width}
\end{figure*}

The learnt energy depends on two variables: the time $t$, and the configuration $\bm{x}$. We probe its behaviour by constructing a two-dimensional landscape over $t$ and a collective variable $d(\bm{x}_t, \bm{x}_0)$. Specifically, $d(\bm{x}_t, \bm{x}_0)$ is the standard deviation of atomic distances between the positions at time t, $\bm{x}_t$, and the positions of the finally generated molecule, $\bm{x}_0$.
\begin{equation}
    d(\bm{x}_t, \bm{x}_0) = \mt{std}(\{|\bm{r}_{1,t}-\bm{r}_{1,0}|, ..., |\bm{r}_{N,t}-\bm{r}_{N,0}|\})
\end{equation}
We call $d$ the ``deformation level'' of $\bm{x}_t$ with respect to $\bm{x}_0$. The construction of samples with a specific deformation level is easy and mimics the way samples are constructed during training (see Section~\ref{section:trained_model_methods}).
Each point in the landscape corresponds to a time $t$ and configuration $\bm{x}_t$ with a specific deformation $d(\bm{x}_t, \bm{x}_0)$.

Panel A in Figure~\ref{fig:diffusion_energy_landscape} shows this landscape using benzene as $\bm{x}_0$.
Two striking features emerge.
First is the clear dependence on time.
At the start of generation ($t>10^{0}$), the energy is not sensitive to deformation levels, increasing only for the most distorted configurations.
However, as $t\rightarrow 0$, the energy increases steeply away from the equilibrium geometry.
This shows that the model has learnt to first act as a simple restorative potential before becoming more chemical near the end. 
Second is the surprising smoothness of the landscape, best seen in Panel B (top), showing the energy along a real diffusion trajectory.
The learnt energy smoothly decreases as the ``soup of atoms'' becomes a molecule (the small energy fluctuations are the result of using a stochastic sampler, not part of the landscape; a deterministic trajectory, grey line, is completely smooth).
Compared to $E_{\text{linear}}$ evaluated on the same trajectory (Panel B bottom), the largest differences appear at intermediate $t$ ($10^{0}<t<10^{-1}$).
Here, atoms often overlap or have unusual valences, causing the \gls{qm} energy to diverge.
In contrast, the diffusion model can handle such irregular configurations while the energy smoothly decreases throughout the generation.

More importantly, the diffusion model has learnt to penalise fragmentation. To show this, we generate molecules by minimising $E_\mt{QM}$ or $E_{\mt{linear}}$ of randomly initialised configurations. In both cases, we first remove any overlaps by relaxing with a Morse potential~\cite{morseDiatomicMoleculesAccording1929}.
Panel C shows that the resulting structures are fragmented, with water and hydrogen being particularly frequent ``side products''.
The high bond dissociation energies make the formation of these small molecules irreversible, forcing the remaining atoms into extremely unsaturated and high-energy structures.
This highlights the importance of the alchemical force $\bm{s}_{\bm{z}}(\bm{x};t)$: It allows diffusion models to dynamically change the element composition, preventing extremely unsaturated structures and enabling changes to the molecular graph without breaking bonds.

Panel D shows how the \gls{qm} and diffusion model's energies change over the course of relaxation.
Minimising $E_\mt{linear}$ causes the \gls{qm} energy to initially increase due to the restorative part of the potential.
Additionally, it produces larger molecules as the high pressure keeps the atoms together long enough for them to form bonds.
The diffusion model consistently assigns a higher energy to the $E_\mt{QM}$ relaxation trajectories, showing a bias towards larger, i.e. less fragmented, molecules.

\citet{mateLearningInterpolationsBoltzmann2023} recently investigated different interpolation schemes between prior and target densities for flow-based generative models.
They showed that a trainable middle potential is needed to ensure coverage of the different modes in the target density.
Our results indicate that, for molecular generation, energy-based diffusion models learn the key aspects of such a specialised potential without explicit separation of $E(\bm{x};t)$ into distinct components during training.
Key features of this middle potential include smoothness despite distorted configurations; element swapping with the alchemical force; and bias toward training data-like structures.

\subsection{Zero shot generation with similarity kernels and evolutionary algorithms}

Although diffusion models achieve excellent results for 3D molecular generation, they have some inherent limitations:
\begin{itemize}
    \item \textbf{Lack of scalability:} Diffusion models are global, with the score acting on each atom dependent on the entire structure. This makes scaling to larger systems difficult.
    \item \textbf{Limited control:} Conditioning requires specialised training~\cite{igashovEquivariant3DConditionalDiffusion2022}. Users cannot easily guide generation or impose constraints.
    \item \textbf{Low transferability} Diffusion models only generate molecules that \textit{globally} match the training distribution. Applying them to new domains requires collecting large datasets (although the amount of new data required can be reduced by transfer learning~\cite{leNavigatingDesignSpace2023a}).
\end{itemize}

However, we believe that interpolating between an analytical prior and quantum mechanical energy remains a sound idea. Building on it and the insight from the previous section, we propose using a similarity kernel as the middle potential to link the two ends of the generation process, avoiding the need to train a model.

Our approach, named Similarity-based Molecular Generation (SiMGen), combines a time-dependent kernel with an evolutionary algorithm.
The kernel functions on local atomic environments, rendering the entire generation process local.
Yet, the atomic representations come from a pre-trained \gls{mlff}, thus making the generation ``zero-shot'' rather than entirely learning-free.

\subsubsection{Similarity kernels as generative models}

Previously, we used $\bm{x}=\{\bm{r},\bm{z}\}$ to represent whole molecules. We can also represent a molecule as a collection of atomic environments $\bm{x}=\{\bm{\chi}_1, ..., \bm{\chi}_N\}$, where $\bm{\chi}_i$ is a descriptor vector, representing the local environment around the atom $i$. The kernel function $k(\bm{\chi}_i, \bm{\chi}_j)$ is a measure of similarity between $\bm{\chi}_i$ and $\bm{\chi}_j$. By maximising this measure, we can directly steer the generation towards a chemical space defined by a set of reference atomic environments $\mathcal{D}_\mt{ref}=\{\bm{\chi}_1, ..., \bm{\chi}_N\}$. 

The correct choice of kernel and its parameters is essential for this strategy to work. Here we use a time-dependent version of the \gls{rbf} kernel:
\begin{equation} \label{eqn:kernel_equation}
    k(\bm{\chi}_i, \bm{\chi}_j; t) = \exp\left(-\frac{{||\bm{\chi}_i - \bm{\chi}_j||}^2}{2{\sigma(t)}^2}\right),
\end{equation}
The \gls{rbf} kernel is a universal approximator but more importantly for us the kernel width $\sigma(t)$ controls the locality of the kernel \textit{and its gradient}~\cite{deringerGaussianProcessRegression2021a}. When $\sigma(t)$ is small, the kernel and its gradient rapidly decay to zero as $||\bm{\chi}_i - \bm{\chi}_j||$ increases, while when $\sigma(t)$ is large, the decay is much more gradual. This ability to control the locality of the kernel is essential, as we will discuss further on.

Since we want the kernel to measure local similarity, we also need the representations $\bm{\chi}_i$ to be local. While any local scheme, such as ACE~\cite{drautzAtomicClusterExpansion2019} or SOAP~\cite{bartokRepresentingChemicalEnvironments2013}, could be used to generate $\bm{\chi}$, we found that learnt descriptors from a pretrained \gls{ml} potential perform better. Section~\ref{section:mace_features} gives details on the construction of such descriptors.

\begin{figure*}[ht!]
    \centering
    \includegraphics[width=0.9\textwidth]{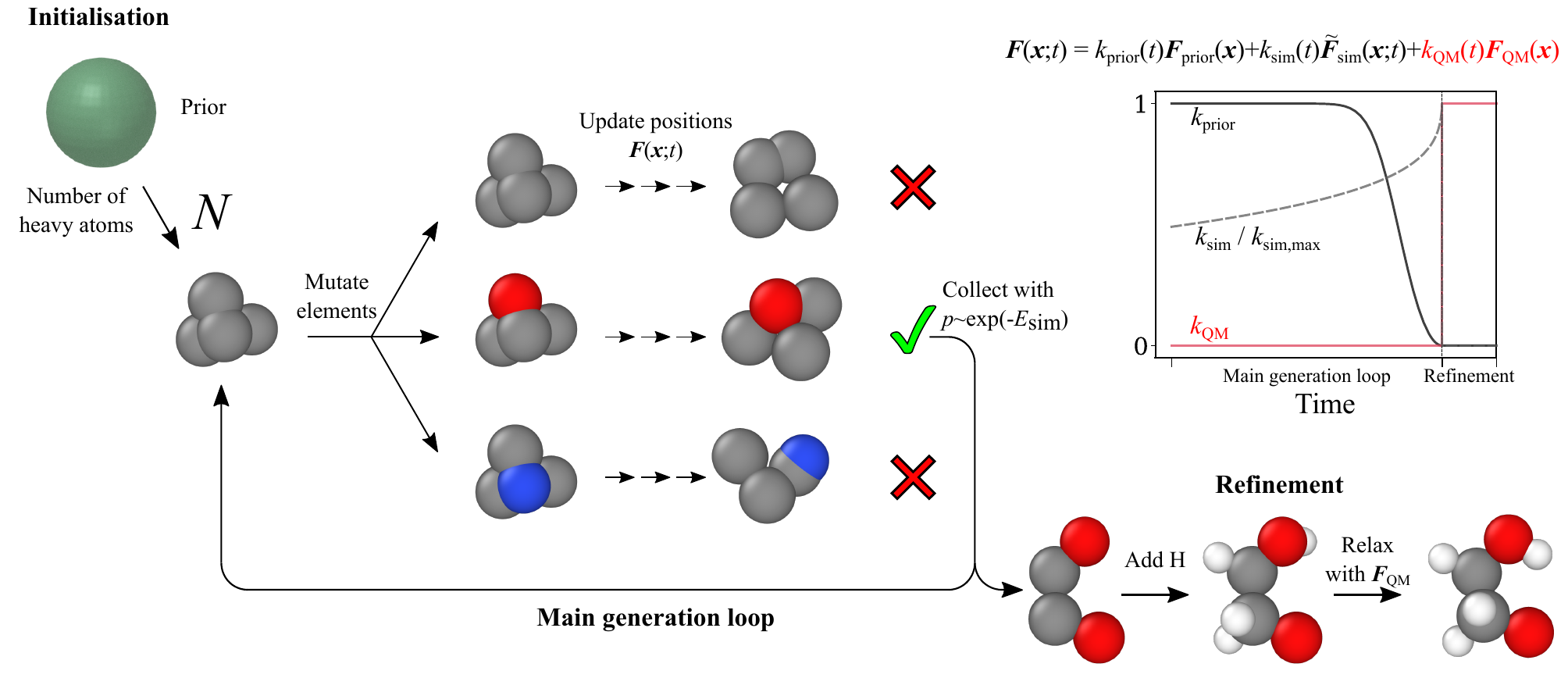}   \caption{\textbf{Molecular generation using similarity kernels} The main loop combines PSO with $\bm{F}_\mt{prior}$ and $\bm{F}_\mt{sim}$ to generate a complete molecule without hydrogen atoms. In the refinement stage, hydrogens are added and a pre-trained \gls{mlff} relaxes the structure.}
    \label{fig:main-scheme}
\end{figure*}

To use the kernel as a generative model, we construct an energy function defined by the reference environments $\mathcal{D}_\mt{ref}$. 
We define the similarity energy of an atom as the negative log likelihood of its environment with respect to the reference data.
\begin{equation}
	E_{\mt{sim},i} =  - \log f(\bm{\chi}_i,t) = - \log \sum\limits_{j\in\mathcal{D}_{\mt{ref}}} k(\bm{\chi}_i, \bm{\chi}_j;t)
\end{equation}
Summing over the atoms yields the total similarity energy of a configuration. 
\begingroup\makeatletter\def\f@size{8.5}\check@mathfonts
\def\maketag@@@#1{\hbox{\m@th\large\normalfont#1}}%
\begin{align}
    E_\mt{sim}(\bm{x};t) &= \sum\limits_{i\in\bm{x}} E_{\mt{sim},i} = - \sum\limits_{i\in\bm{x}} \log \Big( \sum\limits_{j\in\mathcal{D}_{\mt{ref}}} k(\bm{\chi}_i, \bm{\chi}_j;t)\Big) \label{eqn:sim_energy} \\
    \bm{F}_\mt{sim}(\bm{x};t) &= - \nabla_{\bm{x}} E_\mt{sim}(\bm{x};t) = \nabla_{\bm{x}} \sum\limits_{i\in\bm{x}} \log f(\bm{\chi}_i;t). \label{eqn:sim_score}
\end{align}\endgroup
$E_\mt{sim}(\bm{x};t)$ quantifies the similarity of atomic environments in $\bm{x}$ to the reference set $\mathcal{D}_\mt{ref}$ -- the more similar the environments, the lower the energy. The force on an individual atom can be understood as the direction that maximises the local similarity to the reference data. This similarity force generalises the work of \citet{cobelliLocalInversionChemical2022}, allowing the generation of environments that are similar \textit{but distinct} from the reference data.

In state-of-the-art diffusion models, the score function is constructed in such a way that it always points in the direction of the data~\cite{songDenoisingDiffusionImplicit2022, karrasElucidatingDesignSpace2022}. We can emulate this effect by setting a large initial kernel width $\sigma(t)$ and then reducing it as the generation progresses. Expanding Equation~\ref{eqn:sim_score} shows that the similarity force consists of weighted contributions from all reference environments.
\begin{subequations}\label{eqn:similarity_force_expanded_full_equation}
    \begin{align}
         \bm{F}_\mt{sim}(\bm{x};t) &= \frac{1}{\sigma(t)^2}\sum\limits_{i\in\bm{x}}\sum\limits_{j\in\mathcal{D}_{\mt{ref}}} w_j (\bm{\chi}_j - \bm{\chi}_i)^T\nabla_{\bm{r}}\bm{\chi}_i \label{eqn:similarity_force_expanded}\\
         w_j &= \frac{\exp\left(-||\bm{\chi}_i - \bm{\chi}_j||^2 / {(2\sigma(t)^2)}\right)} {\sum\limits_{j\in\mathcal{D}_{\mt{ref}}} \exp\left(-||\bm{\chi}_i - \bm{\chi}_j||^2 / {(2\sigma(t)^2)}\right)} \label{eqn:softmax_weight} \\
         &= \mt{softmax}(-||\bm{\chi}_i - \bm{\chi}_j||^2 / {(2\sigma(t)^2)})
    \end{align}
\end{subequations}
When $\sigma(t)$ is large, all reference environments contribute equally per the softmax, and $\bm{F}_\mt{sim}(\bm{z};t)$ points towards the "mean atomic environment". As $\sigma(t)\rightarrow 0$, only the closest reference environment dominates the force acting on each atom. This transition from large to small $\sigma(t)$ enables atoms to first explore possible arrangements before ultimately settling into specific local minima.

We demonstrate how the width of the kernel affects the similarity energy in Figure~\ref{fig:effect_of_kernel_width}. Each individual plot is constructed by placing an ethanol molecule without hydrogen atoms in the XY plane. The energy at each $(x,y)$ corresponds to $E_\mt{sim}$ if an additional oxygen atom is placed at that coordinate. At the largest $\sigma$ the repulsive terms dominate, as the ``mean atomic environment'' does not contain environments where the atoms overlap. As we decrease $\sigma$, specific minima emerge. Next to the $\beta$-carbon, 4 minima appear. These minima correspond to the formation of either a single bond or a double bond to the new oxygen atom. Only one minimum emerges next to the $\alpha$-carbon. The \ch{O-C-O} fragment can be planar only if one of the oxygen atoms is double bonded to the carbon, making a \ch{O=} site the only possibility. Varying $\sigma(t)$ is crucial for the generation process, as no single value of $\sigma(t)$ is appropriate for the whole generation.

The core of our method is the similarity energy and force $E_\mt{sim}$ and $F_\mt{sim}$. However, to match the correct distributions at the two endpoints, we need to combine the similarity force with a quadratic restorative force early in the generation and the actual quantum mechanically derived force near the end.
\begin{equation}\label{eqn:overall_similarity_score_func}
    \begin{split}
        \bm{F}(\bm{x};t) =  &k_\mt{prior}(t)\bm{F}_\mt{prior}(\bm{x}) 
         + k_\mt{sim}(t)\tilde{\bm{F}}_{\mt{sim}}(\bm{x};t) \\
        & + k_{\mt{QM}}(t)\bm{F}_{\mt{QM}}(\bm{x})
    \end{split}
\end{equation}
$\bm{F}_\mt{prior}(x)$ is the gradient of the log likelihood of the prior distribution. In the case of standard Gaussian prior $\bm{F}_\mt{prior}(x)=-\bm{r}$. In theory, scheduling functions $k_*$ are constrained such that $\bm{F}(\bm{x};T)=\bm{F}_\mt{prior}(\bm{x})$ and $\bm{F}(\bm{x};0)=\bm{F}_\mt{QM}(\bm{x})$. In practice, we use the similarity force from the start, as moves guided solely by the uninformative prior waste computational effort without contributing to the final result. Equation~\ref{eqn:similarity_force_expanded} suggests a natural schedule, $k_{\mt{sim}}(t)  = 1/\sigma(t)^2$, for the similarity force, and we thus rewrite similarity force as the product ${\bm{F}}_{\mt{sim}}(\bm{x};t) = k_{\mt{sim}}(t) \tilde{\bm{F}}_{\mt{sim}}(\bm{x};t)$.  

The force $\bm{F}(\bm{x};t)$ is equivalent to the score component responsible for the atomic positions $s_{\bm{r}}(\bm{x},t)$, yet so far we have largely ignored the question of how to evolve the elemental composition $\bm{z}$. In the next section, we present a method to optimise $\bm{z}$ without having to define and train an alchemical force $s_{\bm{z}}(\bm{x};t)$.

\subsubsection{Handling element swaps}

We can obtain an alchemical force from the similarity energy by differentiating it with respect to the elemental embedding $\bm{z}$.
\begin{equation}
\bm{s}_{\bm{z}}(\bm{x};t)\overset{?}{=} - \nabla_{\bm{z}} E_\mt{sim}(\bm{x};t)
\end{equation}
To represent atomic environments, we use a pretrained model that initially encodes the elements as one-hot categorical vectors. If we used $\bm{s}_{\bm{z}}(\bm{x};t)$ directly, the discrete element encoding would become a continuous vector. While diffusion models can handle both discrete and continuous element embeddings - they are specifically trained to do so - a model trained only on discrete embeddings may fail to generate reasonable atomic environments when applied on structures with continuous element embeddings. To avoid this issue, we instead turn to an evolutionary algorithm.

We use a modified version of \gls{pso}~\cite{kennedyParticleSwarmOptimization1995} to optimise the element composition $\bm{z}$. \gls{pso} explores a wide solution space using a population of particles that share information about the current best solution. We introduce a mutation phase into \gls{pso} for element swapping. Each round begins by creating copies of the current best solution with lowest $E_\mt{sim}$. We then mutate a fraction of atoms in each particle by changing their element. Atoms with the lowest local similarity have the highest mutation probability. Additionally, one particle remains unchanged to allow the possibility that the original $\bm{z}$ is already optimal. After mutation, the particles evolve independently according to Equation~\ref{eqn:overall_similarity_score_func}. This scheme enables element swaps without requiring an explicit $s_{\bm{z}}$ and helps to find overall lower energy solutions by exploring a wider configuration space.

The modified \gls{pso} scheme has a downside -- it struggles with single-valence elements. If we allow swaps to group 1 or 17 elements, the generation process almost always converges to isolated gas molecules. Since these elements have only one nearest neighbour, minimising $E_\mt{sim}$ is trivial by switching to, for example, hydrogen and forming molecular \ch{H2}. To prevent this behaviour, we remove group 1 and 17 elements from the reference data and forbid swaps to these elements during \gls{pso}. We then add back hydrogen atoms separately after the \gls{pso} loop has finished. 

Generation with explicit hydrogen atoms is an ongoing problem in the field. Including explicit hydrogen atoms generally reduces the performance of trained models~\cite{huaMUDiffUnifiedDiffusion2023, vignacMiDiMixedGraph2023} and increases cost without fundamentally improving the quality of the generated structures. As such, many groups choose to treat the addition of hydrogen as a postprocessing task~\cite{joScorebasedGenerativeModeling2022, zangMoFlowInvertibleFlow2020, liuGraphEBMMolecularGraph2021}.

\subsubsection{Full procedure}

\begin{figure*}[ht!]
    \centering
    \includegraphics[width=0.95\textwidth]{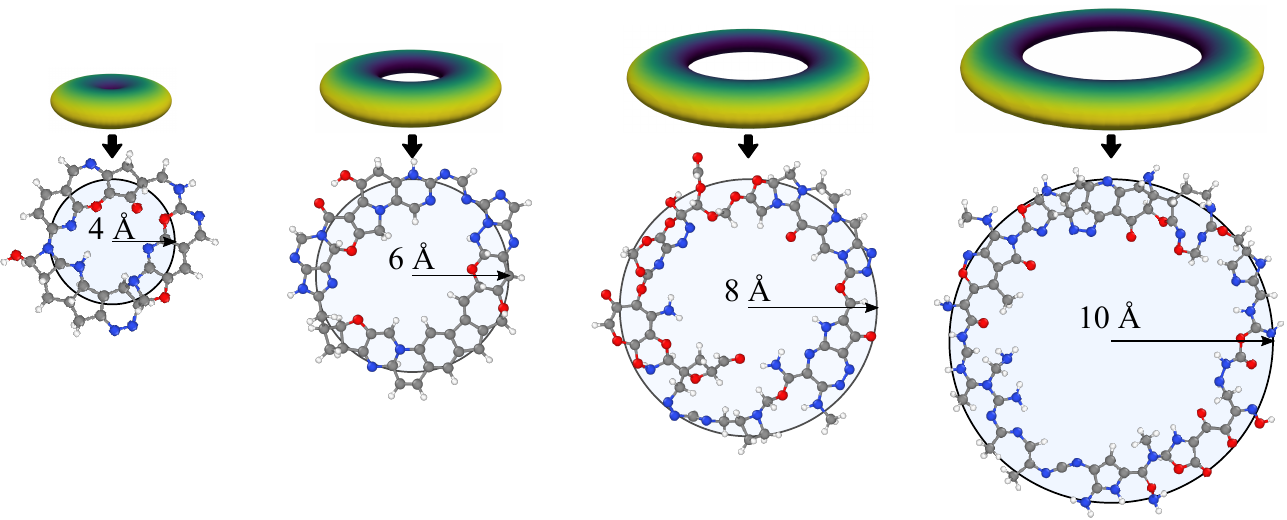}
    \caption{\textbf{Point cloud prior enables generation in an arbitrary shape} Macrocycle-like structures can be generated by combining a circular prior with a flat covariance.} 
    \label{fig:macrocycle-example}
\end{figure*}

Figure~\ref{fig:main-scheme} illustrates the full generation procedure. The main generation loop uses the modified \gls{pso} and the force given by Equation~\ref{eqn:overall_similarity_score_func} to generate a complete structure without hydrogen atoms. Since no hydrogen atoms are present, we set $k_\mt{QM}=0$ throughout this stage (the intermediate configurations without H appear extremely unsaturated, making physical forces less useful). After the main loop, we add back the hydrogen atoms and perform one last element swapping round, now using QM energies to determine the final composition. As a final step, we relax the geometry using physical forces only, $k_\mt{prior} = k_\mt{sim} = 0$.

The final structure of the generated molecule is primarily determined by the main generation loop. The refinement stage adds explicit hydrogen atoms and relaxes the molecule typically without further changes to the heavy atoms. Although we use purpose-trained models for refinement (see Section~\ref{section:refinement}, the generative part of SiMGen does not require specialised models.

To evaluate our method, we generated 1,000 structures using local environments extracted from 256 QM9 molecules~\cite{ramakrishnanQuantumChemistryStructures2014} as reference data. We picked the reference molecules by first sampling two molecules uniformly based on the number of heavy atoms, and the remaining molecules were selected randomly.

We used two criteria to assess the quality of the generated structures: (1) similarity to the reference environments, quantified by $E_\mt{sim}$, and (2) energy evaluated using an accurate force field model. For comparison, we also assessed 1,000 structures produced by the linear interpolation baseline (Section \ref{section:trained-model-and-airss}), as well as 10,000 additional QM9 molecules.

\begin{figure}[p]
    \centering
    \includegraphics[width=0.9\linewidth]{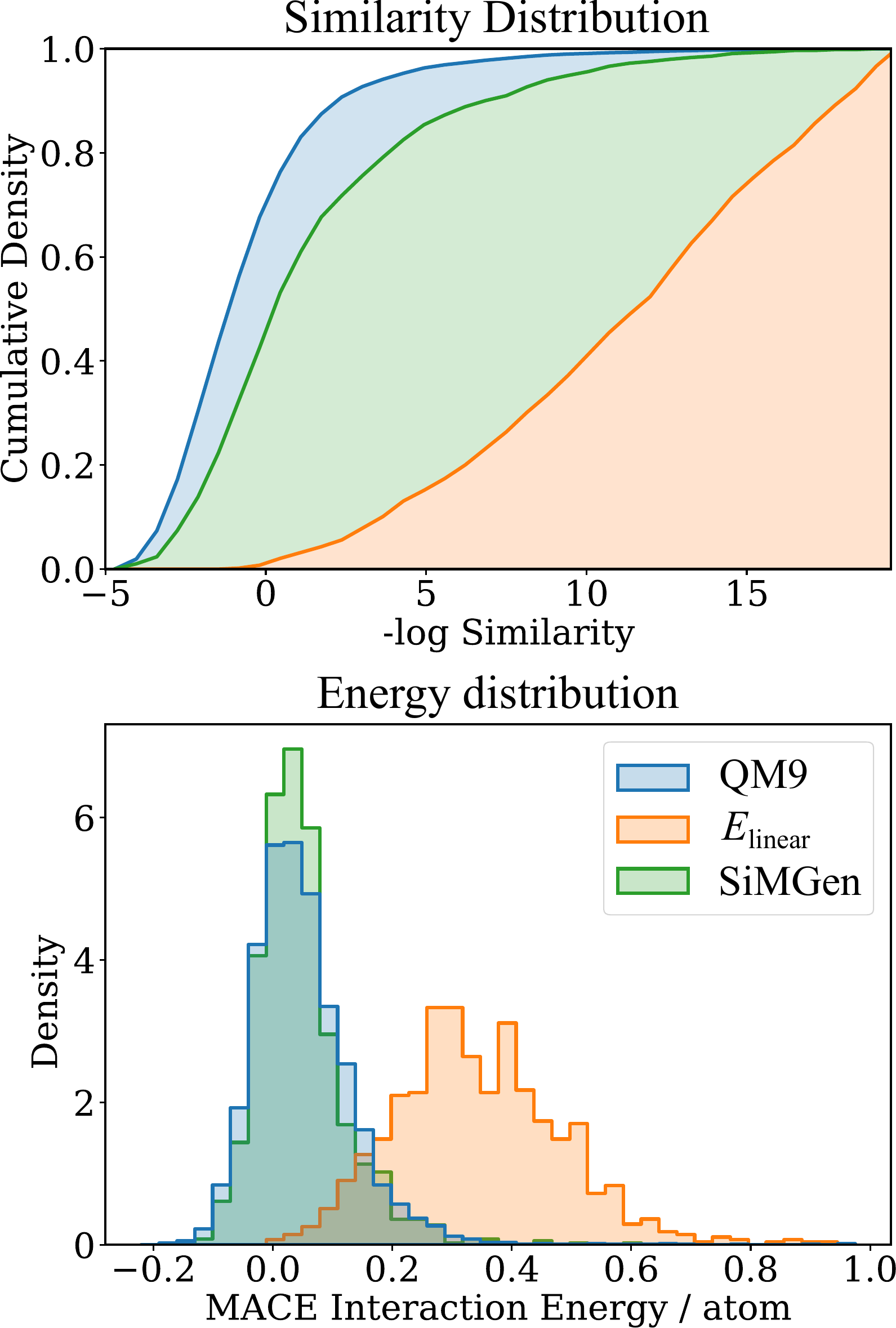}
    \caption{\textbf{Molecules generated with the similarity kernel correctly sample the reference distribution} Molecules generated with the similarity kernel match the reference data both in energy and similarity. The $E_\mt{linear}$ baseline often generates fragmented and highly unsatured structures, resulting in high energy and low similarity. All structures were relaxed with a pretrained \gls{mlff} before comparison.}
    \label{fig:energy_and_similarity_distributions}
\end{figure}

Figure~\ref{fig:energy_and_similarity_distributions} shows the cumulative similarity distribution and the energy distribution for the three test sets. Molecules generated with the similarity kernel have almost identical similarity and energy distributions to molecules from QM9. This suggests our method successfully generates structures that match the reference distribution. In contrast, molecules from the linear scheme have higher energies and are less similar to the reference data. This is unsurprising, as most runs result in fragmented and highly unsaturated structures.

\begin{figure}[p]
    \centering
    \includegraphics[width=0.9\linewidth]{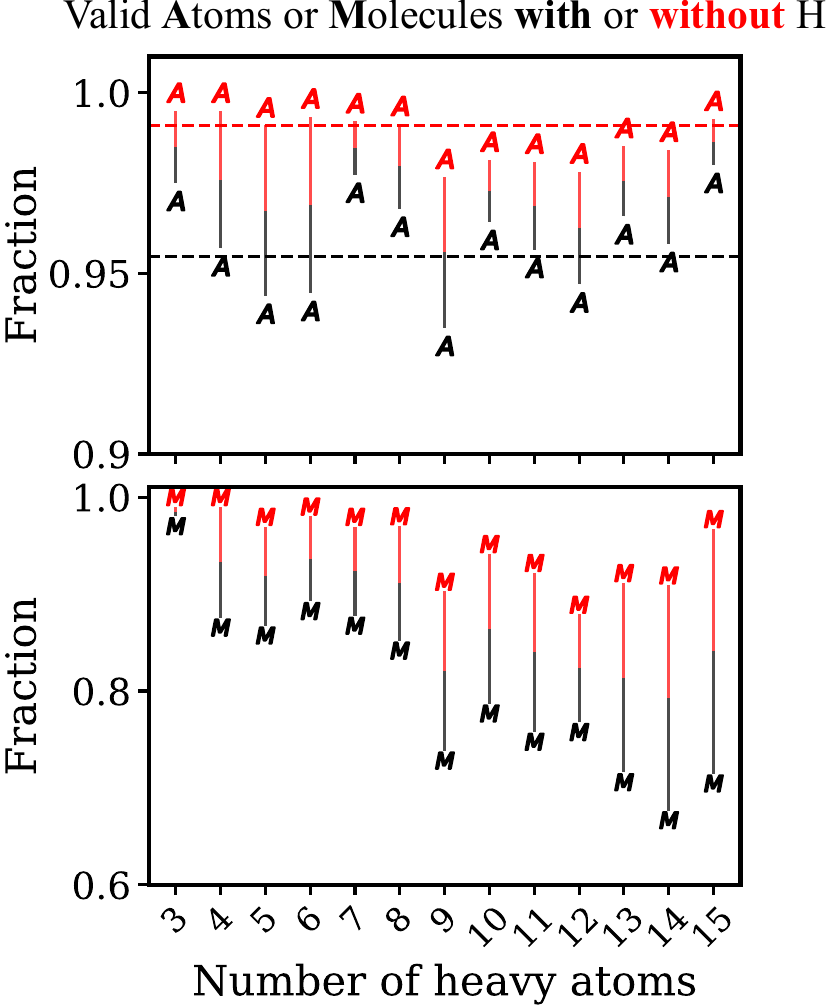}
    \caption{\textbf{Atom validity is independent of molecule size} The plots show fraction of valid atoms or molecules as a function of heavy atoms in the generation. Valid atoms are defined as atoms whose number of closest neighbours $\leq$ their natural valence. A molecule is valid if all its atoms are valid. Before the hydrogenation step, atom validity is 99\%, adding H atoms drops it to 95\%.}
    \label{fig:validity_vs_size}
\end{figure}

Finally, we analyse how the performance of SiMGen depends on the size of the generated molecule. Figure~\ref{fig:validity_vs_size} illustrates the fraction of valid atoms and molecules versus the number of heavy atoms at the start of generation. As expected from a local generator, the percentage of valid atoms remains roughly constant at 99\% before hydrogen addition and 95\% after, regardless of molecule size. However, the fraction of valid molecules (where all atoms are valid) after hydrogenation reveals that the hydrogen addition step is the main failure point of the whole generation pipeline.

\subsection{Shape control}\label{section:shape-control}

Due to training constraints, diffusion models use the standard Gaussian $N(\bm{\mu}=\bm{0}, \bm{\Sigma}=\bm{I}\sigma)$ as the prior for generation.
Since SiMGen does not require training, we can start generation from any distribution.
We take advantage of this flexibility to extend the generation to new priors, allowing full control over the final shape of the generated molecule.

\subsubsection{Anisotropic Gaussian prior}

We start by extending the standard Gaussian to the anisotropic case with diagonal covariance $N(\bm{r};\bm{0},\bm{\Sigma})$. The probability density for a multivariate Gaussian is:
\begin{equation}\label{eqn:gauss_density}
    N(\bm{r};\bm{0},\bm{\Sigma}) \sim \exp \left(-\frac{1}{2}\bm{r}^\mt{T}\bm{\Sigma}^{-1}\bm{r}\right) 
\end{equation}
Taking the energy perspective of probability, we get $E_\mt{prior} = \bm{r}^\mt{T}\bm{\Sigma}^{-1}\bm{r} / 2$ and the force $\bm{F}_\mt{prior} = -\nabla_{\bm{r}} E_{\mt{prior}} = -{\bm{\Sigma}^{-1}}\bm{r}$. To use this in generation, we initialise atom positions from $N(\bm{x};\bm{0},\bm{\Sigma})$ and apply $\bm{F}_\mt{prior}$ in equation~\ref{eqn:overall_similarity_score_func}. 

\begin{figure}[htp]
    \centering
    \includegraphics[width=\linewidth]{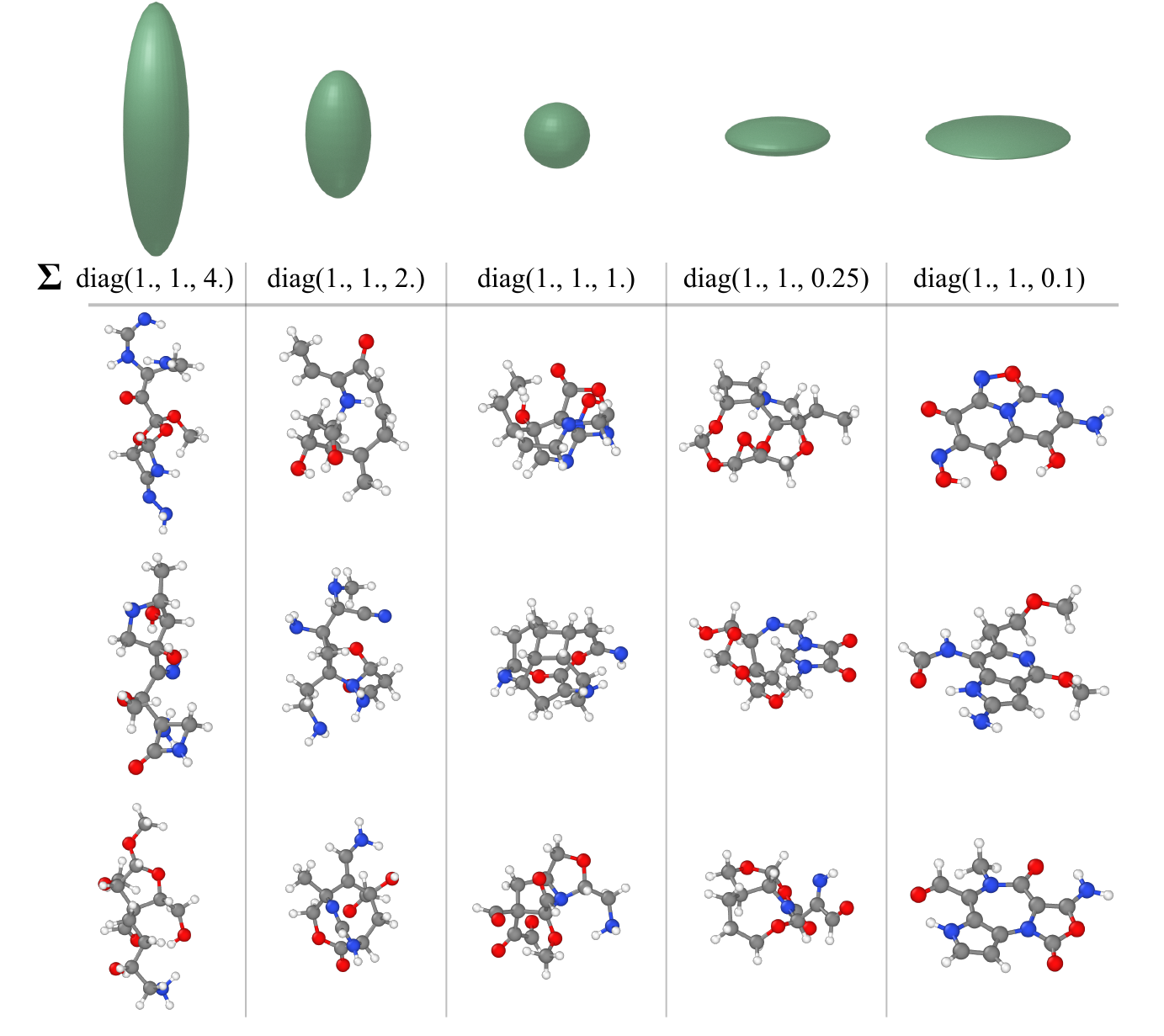}
    \caption{\textbf{The covariance matrix controls the shape of generated structures}}
    \label{fig:multivariate_gauss_prior}
\end{figure}

Figure~\ref{fig:multivariate_gauss_prior} reveals that varying the covariance, $\bm{\Sigma}$, has a dramatic impact on the shape of the generated molecules. Elongated priors create molecules with long aliphatic chains, whereas flattened priors yield conjugated, planar structures.
 
\subsubsection{Point cloud prior}

We can construct a prior with an arbitrary shape via a point cloud. Consider a collection of $N$ Gaussians with means $\{\bm{\mu}_1,\dots,\bm{\mu}_N\}$ and shared covariance $\bm{\Sigma}$. To sample this distribution, we first randomly select a point from the point cloud and then sample $N(\bm{\mu}_i,\bm{\Sigma})$. The density associated with this prior is given by Equation~\ref{eqn:cloud_prior_density}.
\begin{equation} \label{eqn:cloud_prior_density}
    f(\bm{x})=\sum\limits_i^N \exp \left(-\frac{1}{2}(\bm{x}-\bm{\mu}_i)^\mt{T}\bm{\Sigma}^{-1}(\bm{x}-\bm{\mu}_i)\right)
\end{equation}
The energy for this density is $E=-\log f(\bm{x})$. Fortunately, we can take the derivative of this energy analytically with the resulting force given by equation~\ref{eqn:manifold_force}.
\begin{subequations}\label{eqn:point_cloud_prior}
    \begin{align}
        \bm{F}_\mt{prior}&=-\sum\limits_i^N w_i {\bm{\Sigma}^{-1}}(\bm{x}-\bm{\mu}_i),\label{eqn:manifold_force}\\
        w_i &= \frac{\exp \left(-\frac{1}{2}(\bm{x}-\bm{\mu}_i)^\mt{T}\bm{\Sigma}^{-1}(\bm{x}-\bm{\mu}_i)\right)}{\sum_i \exp \left(-\frac{1}{2}(\bm{x}-\bm{\mu}_i)^\mt{T}\bm{\Sigma}^{-1}(\bm{x}-\bm{\mu}_i)\right)}.\label{eqn:cloud_prior_weight}
    \end{align}
\end{subequations}
In practice we use $$w_i = \exp(-|\bm{x}-\bm{\mu}_i|) / \sum_i \exp(-|\bm{x}-\bm{\mu}_i|)$$ for computational simplicity and have not observed any adverse effects on the generation procedure. Note that this derivation is analogous to equations for the similarity force.

In Figure~\ref{fig:macrocycle-example} we combine the point cloud prior with a flattened multivariate Gaussian to generate a sequence of macrocycles with increasing radii. The kernel's reference data contain environments from structures only up to nine heavy atoms, whereas the generated macrocycles range from 45 to 111 heavy atoms. Due to the locality of the kernel, we are able to generate structures much larger than those in the reference data. Although a local builder will eventually make a mistake when generating a very large molecule, these results showcase how a local model could be the backbone for generating large complex structures.

\subsubsection{Interactive generation with ZnDraw}

Some of the most exciting applications of generative modelling in chemistry require constrained generation. For example, to generate a new ligand for a protein or to link two fragments, the generation process must account for the already existing atoms in the structure. We can condition our method by adding atoms or complete fragments into the simulation box and keeping them stationary throughout the generation. While maximising the local similarity, the fragments are then naturally incorporated into the final structure. However, since our method is local, stationary atoms will only affect the generation if they are within the receptive fields of the atoms being optimised by the kernel. Constrained generation thus works best in tandem with a point cloud prior that directs the generation toward the preplaced fragments.

\begin{figure*}[htp!]
    \centering    
    \includegraphics[width=1.0\linewidth]{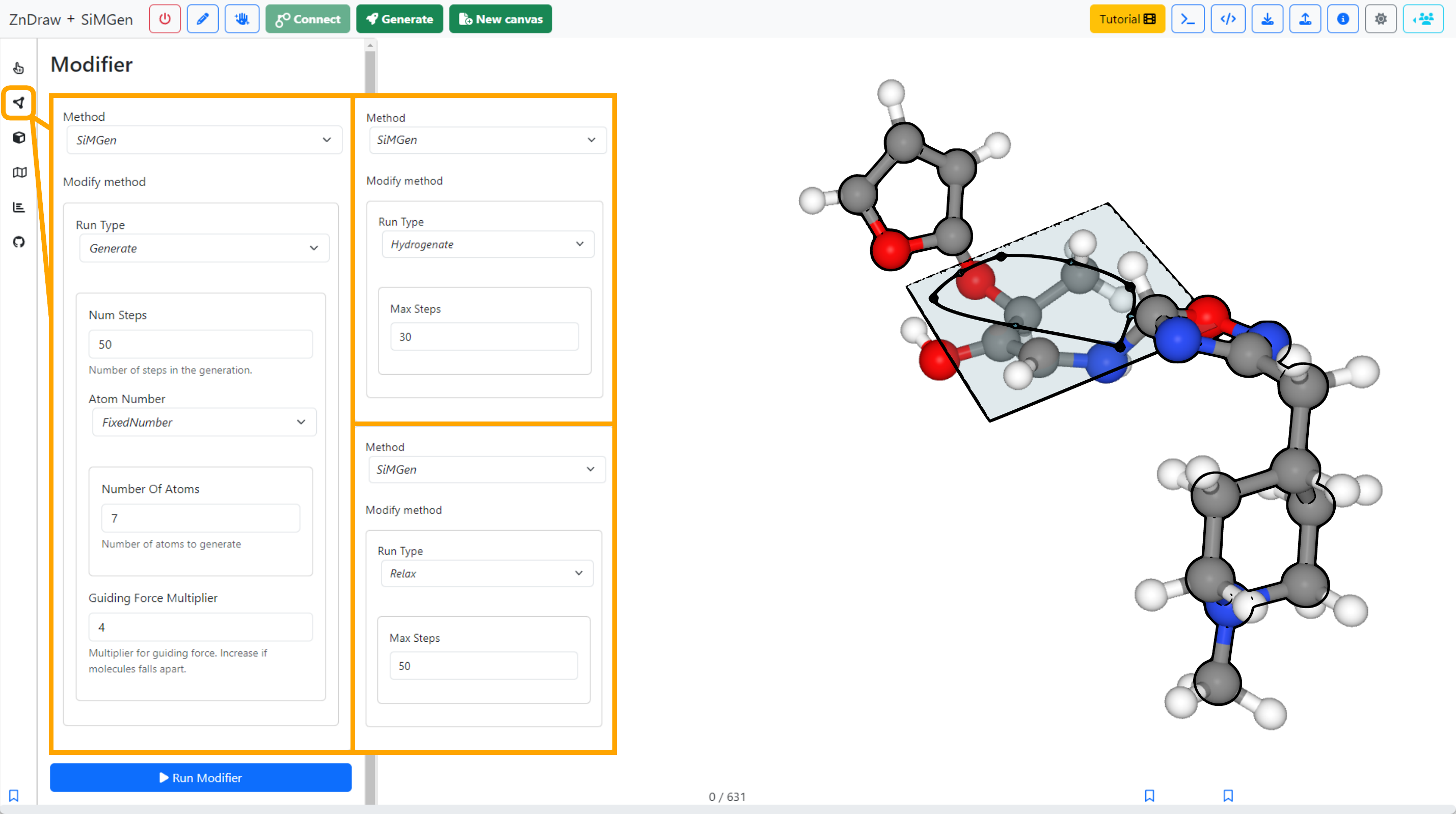}
    \caption{\textbf{Graphical user interface of the ZnDraw package} The menus, highlighted in orange, allow user to adjust the number of generated atoms, strength of the attraction to the prior, and number of steps in the generation. The black outline on the molecule shows the original atom positions before generation. The points on the plane define the guiding point cloud for the generation.}
    \label{fig:zndraw-menus}
\end{figure*}

\begin{figure}[htp!]
    \centering
    \includegraphics[width=0.8\linewidth]{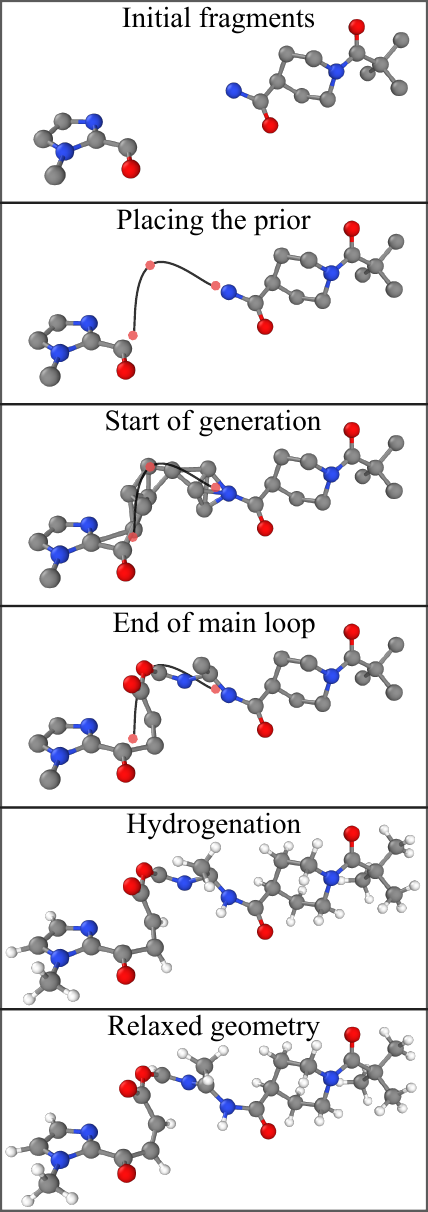}
    \caption{\textbf{Interactive generation with ZnDraw} From top to bottom: the different stages of generation using the interactive tool. The black silhouette shows original atom positions.} 
    \label{fig:interactive-example}
\end{figure}

To enable interactive constrained generation, we developed the ZnDraw software package~\cite{zndrawZenodo}.
ZnDraw provides functionality for visualising, modifying, and analysing atomistic systems.
For generative modelling, users can load structures into ZnDraw, interactively specify 3D point cloud priors, and perform generation driven by these priors.
ZnDraw also includes a graphical user interface to specify generation parameters, as shown in Figure~\ref{fig:zndraw-menus}.
The flexibility to create custom prior shapes makes ZnDraw well-suited for users to design priors tailored to their specific applications.
As an example, Figure~\ref{fig:interactive-example} demonstrates ZnDraw's interactive workflow to link two molecular fragments sourced from~\cite{igashovEquivariant3DConditionalDiffusion2022}.
We provide an online version of ZnDraw with working generative modelling at \url{https://zndraw.icp.uni-stuttgart.de}.

\section{Methods}

\subsection{Similarity kernel details}

The following sections give more information on the main components of the generation and clarify the use of trained models. Briefly, we use a pretrained MACE model to generate representations for the kernel and, during refinement, to swap elements and relax the geometry. In addition, we use a trained model for the hydrogenation step; however, this is a design choice, not a requirement. The hydrogenation model could be replaced with any standard method, e.g. using bond lengths to infer missing hydrogen atoms.

\subsubsection{MACE features}\label{section:mace_features}

We extract the features from a MACE~\cite{batatiaMACEHigherOrder2022, kovacsMACEOFF23TransferableMachine2023} model trained on the SPICE dataset~\cite{eastmanSPICEDatasetDruglike2023} containing 1 million molecules of 3 to 100 atoms. The MACE model is a many-body equivariant message passing neural network (MPNN). We summarize the key steps of the construction of the MACE descriptors.

The first step in MACE is to construct the local neighborhood of an atom based on a cutoff distance, $\mathcal{N}(i) = \{j| r_{ij} \leq r_{\text{cut}} \}$. Then two body information is encoded in a one particle basis as a product of radial basis function $R$, spherical harmonics $Y_{l}^{m}$ and node features $h_{j}$, displayed in Eq. \ref{eq:phi-basis-t} where $\hat{\bm{r}}_{ij}$ denotes the relative positions. The node features are initialized as one hot chemical element, $h_{j}^{(0)} = z_{j}$. The one particle basis is summed over the neighborhood to achieve permutation invariance in Eq.~\ref{eq:atomic-basis-t}, after which $(\nu +1)$-body features are formed in \ref{eq:symmbasis_L1} by taking the tensor product of the atomic basis $A_{i,kl_{3}m_{3}}^{(s)}$ with itself $\nu$ times. The tensor product is symmetrised by contracting the Clebsch Gordan coefficients $\mathcal{C}^{LM}_{\eta, \bm l \bm m}$ where $\eta$ enumerates all possible symmetric couplings to form the symmetrized basis ${\bm B}^{(s)}_{i,\eta k LM}$. A message is formed as a learnable combination of the symmetrised basis in Eq.~\ref{eq:message}. The next node features are for the step $(s+1)$ formed by applying a linear update function on the message in \ref{eq:update}. This operation is repeated $S$ time, always reusing previously constructed node features.
\begingroup\makeatletter\def\f@size{9}\check@mathfonts
\def\maketag@@@#1{\hbox{\m@th\large\normalfont#1}}%
\begin{align}
\label{eq:phi-basis-t}
  \phi_{ij,k \eta_{1} l_{3}m_{3}}^{(s)} &= 
    \sum_{l_1l_2m_1m_2} C_{\eta_1,l_1m_1l_2m_2}^{l_3m_3}
      R_{k \eta_{1} l_{1}l_{2} l_{3}}^{(s)}(r_{ij}) \,\,\times \notag\\
      & \qquad\qquad \times Y^{m_{1}}_{l_{1}} (\boldsymbol{\hat{r}}_{ij}) \bar{h}^{(s)}_{j,kl_2m_2} \\
\label{eq:atomic-basis-t}
    A_{i,kl_{3}m_{3}}^{(s)} &= \sum_{\tilde{k}, \eta_{1}} w_{k \tilde{k} \eta_{1}}
    \sum_{j \in \mathcal{N}(i)}  \phi_{ij,\tilde{k}\eta_{1}l_{3}m_{3}}^{(s)} \\
\label{eq:symmbasis_L1}
  {\bm B}^{(s)}_{i,\eta k LM}
  &= \sum_{{\bm l}{\bm m}} \mathcal{C}^{LM}_{\eta, \bm l \bm m} \prod_{\xi = 1}^{\nu} A_{i,\tilde{k} l_\xi  m_\xi}^{(s)} \\
 \vbox to 20pt{}
 \label{eq:message}
  m_{i,k LM}^{(s)} &=  \sum_{\nu}\sum_{\eta_{\nu}} W_{z_{i} \eta_{\nu} k L}^{(s),\nu} {\bm B}^{(s),\nu}_{i,\eta_{\nu} k LM}\\
 \vbox to 20pt{}
 \label{eq:update}
  h^{(s+1)}_{i,k LM}
    &= \sum_{\tilde{k}} W_{k L,\tilde{k}}^{(s)} m_{i,\tilde{k}LM}^{(s)}
  + \sum_{\tilde{k}} W_{kz_{i} L,\tilde{k}}^{(s)}
  h^{(s)}_{i,\tilde{k}LM} 
\end{align}\endgroup

Experiments with the trained diffusion model (Section~\ref{section:trained-model-and-airss}) show that being able to handle overlapping atoms is important to the generation process. In general, MACE features vary smoothly with changes in the atomic coordinates; however, this is less true when the interatomic distances approach zero. The reason for this is two-fold: one, there is little training data in this region, and two, the features must encode the rapidly increasing repulsive forces as atoms start to overlap. We can avoid this by introducing a distance transform in equation~\ref{eq:phi-basis-t}.
\begin{align}
    \phi_{ij,k \eta_{1} l_{3}m_{3}}^{(s)} &= 
    \sum_{l_1l_2m_1m_2} C_{\eta_1,l_1m_1l_2m_2}^{l_3m_3}
      R_{k \eta_{1} l_{1}l_{2} l_{3}}^{(s)}({\color{red} \tilde r_{ij}}) \,\,\times \notag\\
      & \qquad\qquad \times Y^{m_{1}}_{l_{1}} (\boldsymbol{\hat{r}}_{ij}) \bar{h}^{(s)}_{j,kl_2m_2} \tag{19a} \\
      {\color{red} \tilde r_{ij}} &= r_\mt{min}+\mt{max}(0,r_{ij}-r_\mt{min}) \tag{19b}
\end{align}
Here we used $r_\mt{min}=0.75$~\AA. This distance transform ensures the MACE features remain well-behaved even when atoms are fully overlapping, which occurs frequently at the start of the generation.

The pretrained MACE model used here has two layers, 96 channels (number of $k$), a maximum angular resolution of $l_{max}=3$ (maximum $l_{3}$ in the equation~\ref{eq:atomic-basis-t}), and message equivariance of $L=0$ and correlation of $\nu = 3$ at each layer. For the kernel, we extract the invariant scalar node features of the first layer $h^{(0)}_{i, k00}$. They are many-body descriptors of the local environments of the atom $i$ and contain information within a 5.0 \AA{} around each atom.

The MACE architecture ensures that the extracted features follow the necessary symmetries. Moreover, the obtained representations are differentiable, which is a requirement when using the kernel as a generative model. 

\subsubsection{Generation details}

The main generation loop combines the forces in Equation~\ref{eqn:overall_similarity_score_func}, a modified Heun sampler, and \gls{pso} to generate structures.

\paragraph*{The force} schedule used during generation is given by Equation~\ref{eqn:schedules}. As stated in the main text, the \gls{qm} force is only used for refinement.
\begin{subequations}\label{eqn:schedules}
    \begin{align}
        k_\mt{prior}(t) &= \tanh{(20t^2)}\\
        k_\mt{sim}(t) &= \frac{1}{\sigma(t)^2}=  119(1-(t/10)^{1/4})+1
    \end{align}
\end{subequations}
Although these specific schedules worked well, in general any functions that provide the right balance between the two forces and significantly decrease the kernel width by the end of generation could be effective.

In addition to the schedule, $\bm{F}_\mt{prior}$ requires a multiplier dependent on the size of the generated structure. $|\bm{F}_\mt{prior}|$ effectively creates a volume in which atoms can move freely. Without atom number dependence, the restorative force is either too strong, squeezing many atoms into too small a volume, or too weak, allowing atoms to dissociate. Therefore, $F_\mt{prior}$ is multiplied by a factor $\propto 1/n$ during generation, ensuring that the magnitude of the restorative force matches the size of the molecule.

Although $E_\mt{sim}$ is generally well behaved, its repulsive regions can contain small minima, leading to atom overlap during generation. To avoid this, we add a soft short-ranged repulsive term to $E_\mt{sim}$:
\begin{equation}
    E_\mt{repulsive} = \frac{1}{2}\sum\limits_{ij}\exp(-\alpha r_{ij})
\end{equation}
$\alpha$ is a hyperparameter controlling the range of the repulsion.

\paragraph*{The sampler} used with the similarity kernel builds off Algorithm~2 from reference~\cite{karrasElucidatingDesignSpace2022}. We make two modifications. First, we set a minimum noise level added to the positions each sampler step. Second, we use a custom time step schedule that is linear for the first $N/2$ steps, then decays exponentially for remaining steps. 

\paragraph*{In \gls{pso}}, $N_\mt{particles}$ copies of the current configuration are created. Of these, $N_\mt{particles}-1$ are mutated by swapping elements, while one remains unchanged. Each mutation changes the element of $\ceil{0.2\cdot n_\mt{atoms}}$ atoms, chosen based on their local similarity to the reference set:
\begin{equation}
p_\mt{mutate}(i) \propto \exp(\beta \log f(\bm{\chi}i;t))
\end{equation}
Here $\beta \propto \exp(-t)$ is an inverse temperature. This temperature annealing encourages exploration of diverse compositions in the initial stages of generation and focuses the swapping on the lowest similarity environments at the end of generation. For this work, the element swaps are limited to C, N, and O. After mutation, the particles evolve independently for $n_\mt{freq}$ sampler steps. The next round of \gls{pso} begins by selecting one particle with $p \propto \exp(-\beta E_\mt{sim})$, keeping it as the starting point for the next mutations. For the examples here, we used $N_\mt{particles}=10$ and $n_\mt{freq}=2$.

\subsubsection{Refinement details}\label{section:refinement}

The refinement stage has three steps:
\begin{enumerate}[itemsep=0pt, partopsep=0pt, topsep=2pt]
    \item Addition of hydrogen atoms
    \item Element correction
    \item Relaxation
\end{enumerate}

\paragraph*{Hydrogenation} Typically, when generating without hydrogen, generative models use bond lengths to infer bond orders and, in turn, the number of hydrogens required to satisfy the valence of each atom~\cite{joScorebasedGenerativeModeling2022, zangMoFlowInvertibleFlow2020, liuGraphEBMMolecularGraph2021, hoogeboomEquivariantDiffusionMolecule2022}.
This approach is also applicable here: We can use the bond lengths to determine how many hydrogens to add and create valid molecules. However, this method fails for molecules with extensive conjugation, such as benzene, where the bond order falls between discrete values.

Instead, we trained a MACE model to predict the number of additional bonds required to satisfy the valence (\url{https://github.com/RokasEl/hydromace}). The predicted number of hydrogen atoms is added to the structure using rejection sampling for the initial positions. Then, the pretrained MACE model is used to correct the positions.

\paragraph*{Element correction} The PSO or the hydrogen addition steps can result in high-energy element combinations or incorrect valences. We correct this through a final element swap using MACE energies.

The element correction involves $n_\mt{swap}$ individual mutation rounds. Each round, we select the site with the highest MACE interaction energy (site energy minus the energy of the isolated atom). An ensemble of mutants is generated consisting of the set of single-element swaps over the selected site and its closest neighbours. For example, for a C-O fragment the set is [C-O, C-N, C-C, N-O, O-O].

Of these mutants, we select the one with the lowest MACE energy to proceed. Once an atom is mutated, it is fixed and will not be selected or mutated in subsequent steps. We use $n_\mt{swap}=n_\mt{heavy\ atoms}$, so that each heavy atom is checked at least once.

\paragraph*{Relaxation} To obtain the final structure, we relax the whole molecule using the LBFGS algorithm with the pretrained MACE force field.

\subsubsection{Shape of QM9 molecules}

In Section~\ref{section:shape-control}, we showed that the choice of prior has a substantial effect on the final shape of generated molecule. As such, to generate QM9-like molecules we need to know what is the average molecular shape.

If $\sigma^2_1 \leq \sigma^2_2 \leq \sigma^2_3$ are the size-ordered variances along the principal axes of a molecule, then the covariance of a Gaussian prior that best fits the shape of the molecule is $\bm{\Sigma}=\mt{diag}(1,\sigma^2_2/\sigma^2_1, \sigma^2_3/\sigma^2_1)$. Note that the absolute values of the variances are not important, since the volume of the prior is automatically scaled with the number of atoms. 

\begin{figure}[h]
    \centering
    \includegraphics[width=0.9\linewidth]{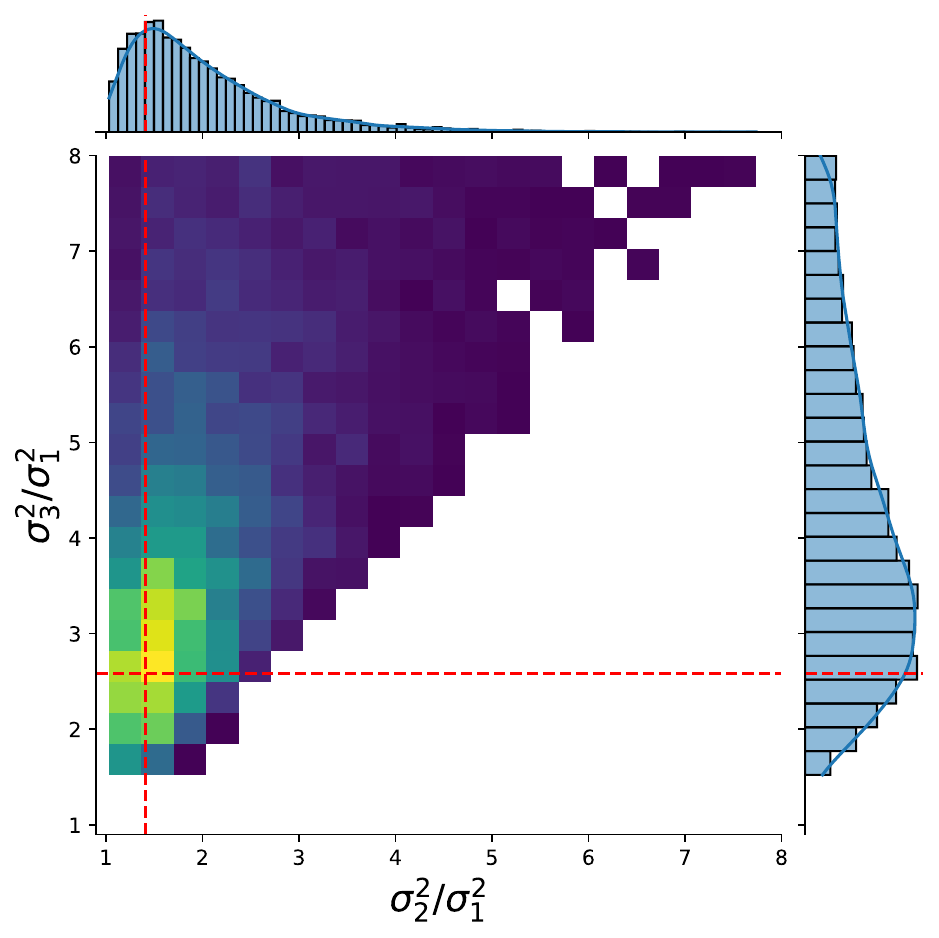}
    \caption{\textbf{Shape of QM9 molecules} The axes correspond to variance ratios along molecular position principal axes. Based on kernel density analysis, the most likely variance ratio is (1.,1.4,2.6) which is also indicated by the red lines.}
    \label{fig:QM9_shape}
\end{figure}

Figure~\ref{fig:QM9_shape} shows the variance ratios for 10,000 randomly selected molecules from the QM9 dataset. Fitting a Gaussian kernel density estimate, we find that the distribution peaks at $(\sigma^2_2/\sigma^2_1, \sigma^2_3/\sigma^2_1)\approx(1.4,2.6)$. 

Thus, the molecular shape in the QM9 dataset is best approximated by a covariance $\bm{\Sigma}=\mt{diag}(1.,1.4, 2.6)$. Whenever QM9-like molecules were generated in the text, we used a prior with this specific covariance.

\subsection{Energy-based diffusion model}~\label{section:trained_model_methods}

The energy-based model uses a modified MACE architecture by adding a time encoding. Specifically, time is positionally encoded~\cite{vaswaniAttentionAllYou2017} and combined with the node features:
\begin{equation}
    \tilde{h}^{(0)}_{i,k} = \mt{MLP}({h}^{(0)}_{i,k}, f(t))
\end{equation}
where ${h}^{(0)}_{i,k}$ are the original MACE features. $\tilde{h}^{(0)}_{i,k}$ replaces ${h}^{(0)}_{i,k}$ in subsequent uses within MACE.

To turn the modified MACE into a generative model, we use the preconditioning proposed by \citet{karrasElucidatingDesignSpace2022}. In this design, the noise schedule is linear $\sigma(t)=t $ and noised samples are created by adding noise without scaling.
\begin{equation}
    \bm{x}_t = \{\bm{r}_t, \bm{z}_t\} = \{\bm{r}_{0}+N(\bm{0}, \bm{I}t^2), \bm{z}_{0}+N(\bm{0}, \bm{I}t^2)\}
\end{equation}

Time $t$ is sampled from a log-normal distribution $\log t \sim N(P_\mt{mean}, P^2_\mt{std})$. Denoised predictions are obtained by wrapping the score in a preconditioning scheme.
\begin{subequations}
    \begin{align}
        D_{\bm{r}}(\bm{x}_t,t) &= c_{\mt{skip}}(t)\bm{r}_{t}+c_{\mt{out}}(t)\nabla_{\bm{r}} E(c_{\mt{in}}(t)\bm{x}_t; c_{\mt{noise}}(t)\cdot t) \\
        D_{\bm{z}}(\bm{x}_t, t) &= c_{\mt{skip}}(t)\bm{z}_{t}+c_{\mt{out}}(t)\nabla_{\bm{z}} E(c_{\mt{in}}(t)\bm{x}_t; c_{\mt{noise}}(t)\cdot t)
    \end{align}
\end{subequations}
The loss is the sum of position and element denoising losses:
\begin{equation}
    L = \frac{1}{\lambda(t)} \left(||\bm{r}_0 - D_{\bm{r}}(\bm{x}_t, \sigma(t))||^2 + ||\bm{z}_0 - D_{\bm{z}}(\bm{x}_t, \sigma(t))||^2\right)
\end{equation}
The exact scaling $c_{\cdot}(t)$ and weighting $\lambda(t)$ functions can be found in Table 1 in reference~\cite{karrasElucidatingDesignSpace2022}. Molecules are generated by initialising from $N(\bm{0}, \bm{I}\sigma(t_\text{max})^2)$ and using the stochastic sampler defined in Algorithm~2 in reference~\cite{karrasElucidatingDesignSpace2022}.

The model used to generate the energy landscape shown in section~\ref{section:trained-model-and-airss} was trained on 80\% of the QM9 dataset for 300 epochs using the one-cycle learning rate policy~\cite{smithSuperConvergenceVeryFast2018}. The model used a cutoff of 10 \AA{} (this corresponds to a global model), 16 radial basis functions, two interaction layers, 64 channels and message equivariance of $L=1$ and correlation of $\nu=3$ at each layer.

For computing the $E_\mt{linear}$ and $E_\mt{QM}$ baselines, we used PM6~\cite{stewartOptimizationParametersSemiempirical2007} with MOPAC~\cite{MOPAC} as a surrogate for the real \gls{qm} energy. 

\subsection{ZnDraw details}

To enable interactive molecular generation, a visualisation of the system, together with the ability to draw 3D point cloud priors, is needed.
Furthermore, a connection between the interface and HPC resources has to be established.
We achieve this in ZnDraw by developing it as a web application that communicates through websockets using the socket.io standard.
The server is built in Python and supports file input through the Atomic Simulation Environment~\cite{larsenAtomicSimulationEnvironment2017a} but can also read files complying with H5MD~\cite{debuylH5MDStructuredEfficient2014}.
The visualisation is realised through the JavaScript package three.js.

A hosted ZnDraw instance supports multi-user access, including sharing sessions for a real-time collaborative experience.
Each session can be connected to one or more Python kernels for manipulating the data.
\begin{lstlisting}[language=Python]
from zndraw import ZnDraw
vis = ZnDraw(url="<url>", token="<token>")
\end{lstlisting}
The connection to the SiMGen software is realised through ZnDraw's plugin interface.
Methods to modify the scene with a given set of parameters can be added using Pydantic, defining the parameters that will be displayed through the user interface.
\begin{lstlisting}[language=Python]
from pydantic import BaseModel

class MyModifier(BaseModel):
    parameter: float = 3.14
    
    def run(self, vis: ZnDraw): ...

vis.register_modifier(MyModifier)
\end{lstlisting}
ZnDraw can be installed locally through \verb|pip install zndraw| on all standard operating systems and uses a command-line interface \verb|zndraw <file>| to visualise molecular structures or interface with a local version of SiMGen.

Furthermore, it is possible to view remote data made available using the ZnTrack~\cite{zillsZnTrackDataCode2024} package, as exemplified in the SiMGen live demo by the reference dataset and the hydrogenation model.

\section{Code availability}

\begin{itemize}
    \item The code for SiMGen is available at \url{https://github.com/RokasEl/simgen}.
    \item The code for ZnDraw is available at \url{https://github.com/zincware/ZnDraw}. The package can be installed via \texttt{pip install zndraw}.
    \item The ZnDraw and SiMGen demo is hosted at \url{https://zndraw.icp.uni-stuttgart.de}.
\end{itemize}

\section{Conclusions}

We have presented SiMGen, a method for generating molecules in 3D using a similarity kernel as the main driving force. The method maximises the local similarity of each atom in the generation and uses an evolutionary algorithm to optimise the element composition. We show that we can successfully sample a reference distribution without training a specialised generative model.

We can initialise the generation from any prior distribution, giving us the ability to control the shape of the generated structure. Combining the shape control with the local nature of the similarity kernel allows us to generate structures with complex shapes and with many more heavy atoms than any structure in the reference data.

To realise an interactive drawing of the prior distribution, we have developed the ZnDraw visualiser, for which we provide an online demo.

\section*{Acknowledgements}
We thank Tam\'as K. Stenczel for initially proposing the idea of using similarity kernels for molecular generation, and Lars L. Schaaf for helpful discussions.

R.E. and I.B. acknowledge support by the University of Cambridge Harding Distinguished Postgraduate Scholars Programme.

S.W.N. acknowledges support from the European Union's Horizon 2020 research and innovation program under the Marie Skłodowska-Curie Actions (Grant Agreement 945357) as part of the DESTINY PhD program, as well as support from the European Union's Horizon 2020 research and innovation program under Grant Agreement 957189 (BIG-MAP).

C.H. and F.Z. acknowledge support by the Deutsche Forschungsgemeinschaft (DFG, German Research Foundation) in the framework of the priority program SPP 2363, “Utilization and Development of Machine Learning for Molecular Applications - Molecular Machine Learning” Project No. 497249646.

C.H. and F.Z. acknowledge further funding though the DFG under Germany's Excellence Strategy - EXC 2075 - 390740016 and the Stuttgart Center for Simulation Science (SimTech).

Access to CSD3 was obtained through a University of Cambridge EPSRC Core Equipment Award EP/X034712/1.

\bibliography{references}

\end{document}